\documentclass[12pt,onecolumn]{IEEEtran}
\usepackage{cite}
\usepackage{graphics}
\usepackage{times}

\begin{document}

\title{Overlay Cognitive Radio in \\ Wireless Mesh Networks}

\author{Ricardo Carvalho Pereira, Richard Demo Souza, \\ and Marcelo Eduardo Pellenz
\thanks{R. C. Pereira and R. D. Souza are with CPGEI, UTFPR. Av. Sete de Setembro, 3165, Curitiba, PR, 80230-901, Brazil. Ph: +55 41 33104686, Fax: +55 41 33104683. rcapereira@timbrasil.com.br,richard@cpgei.cefetpr.br.}
\thanks{M. E. Pellenz is with PPGIA, PUCP-PR. Rua Imaculada Concei\c c\~ao, 1155, Curitiba, PR, 80215-901, Brazil. Ph: +55 41 32711777, Fax: +55 41 32712121. marcelo@ppgia.pucpr.br.}}

\maketitle

\begin{abstract}
In this paper we apply the concept of overlay cognitive radio to the communication between nodes in a wireless mesh network. Based on the overlay cognitive radio model, it is possible to have two concurrent transmissions in a given interference region, where usually only one communication takes place at a given time. We analyze the cases of wireless mesh networks with regular and random topologies. Numerical results show that considerable network capacity gains can be achieved.
\end{abstract}

\section{Introduction}

\PARstart{W}{ireless} Mesh Networks (WMNs) are expected to resolve some of the limitations and to significantly improve the performance of ad hoc networks, wireless local area networks (WLANs), wireless personal area networks (WPANs), and wireless metropolitan area networks (WMANs) \cite{akyildiz.05}. According to its architecture, WMNs may be seen somewhere between a well structured WLAN and an ad hoc network. On one side, each user accesses a fixed gateway in order to establish a communication with the internet. On the other side, every node can reach any other node within a coverage area dictated by the transmission range or be routed through the destination by adjacent users. In WMNs, each user node can operate either as a host or as a wireless router, forwarding packets on behalf of another node that may not be within  the coverage range of a wireless gateway. The coverage extension obtained by allowing the users to hope along other users in order to reach the gateway is an important feature not only to broader the coverage area, but also to allow users to connect from indoor environments or from places with high penetration losses. This results on a substantial reduction of gateways deployment and, consequently, the amount of investments and operational expenses.

The WMN technology presents a set of features that make it an interesting approach for future networks. Some of them are related to the reduced investments required to settle a startup network and to keep it growing, because the technology can be installed incrementally, one node at a time, just as needed. In order to increase the network capacity, accordingly to the demand, more gateways can be added. The reliability is improved too, since the mesh structure ensures the availability of multiple paths for each node in the network. If a node or a gateway fails, the traffic is re-routed to available nodes and gateways. Another interesting characteristic of WMNs is the fact that the network coverage increases with the number of gateways and users \cite{jun.03}. These features are driving the evolving of conventional cellular networks into Multihop Cellular Networks aiming at the enhancement of coverage, data rates, QoS performance in terms of call blocking probability, bit error rate, as well as QoS fairness for different users \cite{le.07}.

Differently of ad hoc networks, the user traffic pattern in WMNs is, essentially, between a user and a gateway \cite{jun.03}. Even though, there is some signaling traffic among users for routing, configuration and control procedures, since this network is supposed to be dynamically self-forming, self-healing and self-organizing \cite{akyildiz.05}. Users can be stationary or mobile. Mobile users are able to be connected or roam as long as they are supported by the WMN coverage.

Usually, the capacity of WMNs is derived based on the capacity of ad hoc networks and it is affected by many factors such as\cite{akyildiz.05}: network architecture, network topology, traffic pattern, network node density, number of channels used for each node, transmission power level, node mobility, to say a few. In \cite{gupta.00} analytical lower and upper bounds of wireless network capacity are determined for the stationary case. The throughput capacity per node reduces significantly when the node density increases, such that it is $O\left(1/{\sqrt n}\right)$, where $n$ is the number of nodes. For a mobile scenario, in \cite{gross.01} it is shown that long term per node throughput can stay constant over the network.

In \cite{jun.03} it is demonstrated that the existence of gateways in WMNs introduces hot spots in the network that act as bottlenecks. Due to the presence of these bottlenecks, the available capacity for each node is reduced to $O\left(1/n\right)$, where $n$ is the number of users for one gateway. Nandiraju {\it et al} \cite{nandi.07} points out that although there are several research solutions to the capacity problem for ad hoc networks, considering the  differences between WMNs and ad hoc networks, many open research issues still exists. For instance, such open research issues include a strategy to decide on the optimal placement of gateways, optimal route selection strategies to increase the throughput of the network, and cross-layer design for network capacity improvement. In \cite{prasad.08}, the concept of layerless communications is introduced, which opens a new perspective to provide reliable and high-quality end-to-end performance in wireless multimedia networks, by means of a global design and optimization. Recently, in \cite{Vu.08} it was shown that in a dense wireless ad hoc network in which nodes may cooperate, a hierarchical cooperation strategy can achieve an asymptotically constant throughput per node.

In this paper we discuss the application of a new communication paradigm to WMN, the cognitive radio model. At first, cognitive radios were viewed as a solution to the problem of overcrowded spectrum by opportunistic communication \cite{mitola.00}. The basic idea of cognitive radio is to explore the existence of room in the licensed spectrum bands to accommodate secondary (unlicensed) wireless devices without disrupting the communications of the primary (licensed) users of the spectrum. The cognitive radio concept can be implemented in different ways, which may try to underlay, interweave, or overlay \cite{jafar.07a,goldsmith.08} the secondary user's signals with those of the primary users, such that the primary users of the spectrum are as unaffected as possible.

We are particularly interested in the overlay approach because of its special feature of allowing concurrent transmissions within a given interference region. In this paper we investigate the effects, in terms of network capacity, of applying the overlay cognitive communication model between nodes in a WMN. Our numerical results show that the proposed scheme can bring considerable capacity gains. Moreover, one of the key challenges of the overlay cognitive radio model, the secondary transmitter non-causal knowledge of the message to be sent by the primary transmitter, is overcame by the sequential nature of WMNs transmissions.

The rest of this paper is structured as follows. In Section II, we briefly discuss WMNs capacity. In Section III we revise some cognitive radio models found in the literature. In Section IV, we apply the overlay cognitive communication model in a WMN. Section V presents numerical results that show the potential capacity improvements of this approach. Section VI concludes the paper.

\section{Capacity of Simple WMNs}
Let us suppose the case of a simple WMN network that consists of a chain of nodes. We assume that only one node in a given region (interference region) transmits at a time (assuming a single frequency channel and no code-division multiple access). The interference region is defined by the union of the interference range of each radio regarding the communication link under consideration. The reason to include the interference range regarding the receiver is based on the fact that this node, receiving a transmission, must transmit back to the transmitter information related to contention procedures, acknowledge and channel state messages, routing procedures and so on. Although a more realistic interference model would be continuous and take some aspects of the wireless environment into account, we follow the discrete interference model used by \cite{li.01}. Such a discrete model is also used in \cite{jun.03}.

\begin{figure}[!tb]
 \begin{center}
\resizebox{60mm}{!}{
\includegraphics{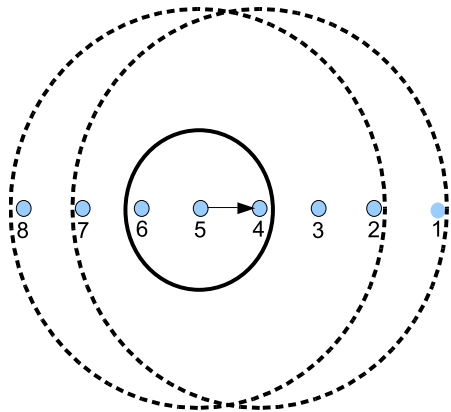}
}
\end{center}
\caption{Example of the discrete transmission and interference range model used in this paper.} \label{fig:fig0}
\end{figure}

An example of the discrete interference model is depicted in Figure \ref{fig:fig0}. The solid line circle is the transmission range of node 5. The dotted line circles are the interference ranges of nodes 5 and 4. Note that in this example we considered the interference range to be three times the transmission range. In this model we assume that if the receiver is closer to the transmitter than the transmission range radius, the receiver will experiment error free reception. Moreover, any radio within the interference range of an active radio is forbidden to establish a communication, since its transmission would affect the active radio. This assumption imposes that during the communication between nodes 5 and 4, nodes 8, 7, 6, 3, 2 and 1 are forbidden to communicate because they are within the interference range composed by the union of the interference ranges of nodes 5 and 4.

Moreover, we assume that  nodes are not able to simultaneously transmit and receive. In other words, they must follow the half-duplex constraint. The case where only the node furthest from the gateway generates traffic to be sent to the gateway was analyzed in \cite{li.01}, while the case of all nodes generating traffic was analyzed in \cite{jun.03}. The analysis we carry out in this paper is based on that presented in \cite{li.01}, since we consider that only one node (or a few nodes) generates traffic. The case of traffic coming from the gateway to the nodes would be analyzed in the same way. The generalization to the case of all nodes generating traffic can be done by analyzing each node separately.

The capacity of the WMN under consideration is a function of the extent of the interference that a node generates to other nodes. For instance, suppose that the transmission range is considered to be the distance of one unit (hop) when transmitting at the maximum power level, and that the interference range is the same. Then, since  nodes can not transmit and receive at the same time, in \cite{li.01} it is shown that the network capacity equals $B/3$, where $B$ is the capacity of the point to point communication between nodes.

\begin{figure}[!tb]
 \begin{center}
\resizebox{120mm}{!}{
\includegraphics{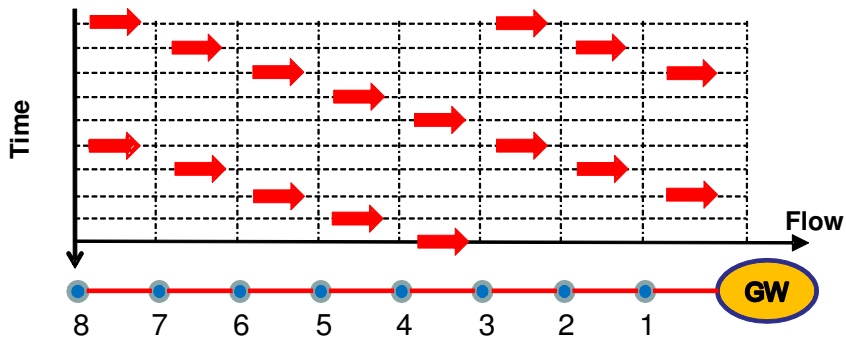}
}
\end{center}
\caption{Example of a WMN composed of a chain of nodes and a gateway. The arrows represent the links which are transmitting at a given time. The interference range is three hops.} \label{fig:fig1}
\end{figure}

If we consider a more realistic scenario, where the interference range is greater than the transmission range\footnote{According to \cite{jun.msc} the interference range is typically more than twice the transmission range.}, then the network capacity is even smaller. Suppose that the interference range is three times the transmission range, as is the case in Figure \ref{fig:fig0}. So, if there is a transmission between two nodes, any node within the interference region, dictated by the interference ranges of both transmitter and receiver, is not allowed to transmit. Figure \ref{fig:fig1} shows an optimum transmission schedule for this network, considering the interference range to be three times the transmission range, as discussed above. In the figure the arrows represent the links which are transmitting at a given time. From the figure we can see that concurrent transmissions occur only for every other fifth link. Therefore, the capacity of the WMN in Figure \ref{fig:fig1}, equals $B/5$. From now on, in this paper, we will consider the interference range to be three times the transmission range.

In a real scenario, using real radios and real protocols, the capacity would be even smaller. However, since in this paper we discuss a theoretical novel transmission scheme for WMNs, which is not based in any current wireless communications standard, we focus in the simplified case of ideal protocols\footnote{A good starting point for designing a real MAC protocol for the method we propose in this paper is the work in \cite{shea}.}.

\section{Cognitive Radio Models}
Cognitive radios are able to listen to the surrounding wireless channel, make decisions on the fly, and encode data using a variety of schemes in order to better explore the channel characteristics and to mitigate the interference \cite{devroye.06a}. As mentioned earlier, the cognitive radio model can be classified into underlay, interweave, or overlay models \cite{goldsmith.08,jafar.07a}.

In the underlay model, the secondary users protect primary users by enforcing a spectral mask on the secondary signal, so that the interference generated by the secondary devices is below the acceptable noise floor for the primary user of the spectrum \cite{jafar.07a}. Due to the spectral mask constraint imposed to the secondary users, this model is suitable for short range applications. Moreover, it is assumed that the secondary cognitive radio has knowledge of the interference caused by its transmitter to all non cognitive users. The interference constraint for the non cognitive users may be met by using multiple antennas to guide the cognitive signals away from the noncognitive receivers, or can be spread below the noise floor, then despread at the cognitive receiver \cite{goldsmith.08}. In \cite{taranto.08}, the authors propose an adaptive antenna array approach, while the latter technique is the basis of both spread and ultrawideband (UWB) communications. Due to the usual restriction of short range applications of this type of cognitive model, it will not be in our discussion on the connection of cognitive radios and WMNs.

The interweave model requires that the secondary transmitter avoids interference to the primary user by transmitting only over spectral segments unoccupied by the primary radios. This approach brings an improvement of the spectrum utilization by opportunistic frequency reuse over the spectrum holes \cite{goldsmith.08}. The interweave technique requires knowledge of the activity information of the primary users of the spectrum.  In \cite{jafar.07b} the authors propose the two-switch interweave model, which enables secondary activity only if there is no activity of the primary users within the interference range of both secondary transmitter and receiver.

Figure \ref{fig:fig2} shows a block diagram representing the two-switch model as proposed in \cite{jafar.07b}. $ST$ is the secondary transmitter and $SR$ is the secondary receiver, while $PU$ is the primary user as seen by the transmitter or the receiver. $X_S$ is the message that $ST$ wants to send to $SR$. $S_t$ and $S_r$ are the secondary transmitter and receiver switches. $S_t$ ($S_r$) assumes the value of 1 when primary user activity is not sensed at the transmitter (receiver). If a primary user is sensed then the corresponding switch assumes the value of 0. $Y_S$ is the message received by $SR$, while $Z_S$ is random noise. In this cognitive radio model the capacity of the primary transmitter is unaffected by the behavior of the secondary transmitter. Let $P_S$ be the secondary transmit power. Based on these definitions, in \cite{jafar.07b} the authors give an upper bound on the capacity for the secondary communication with global side information, $C_{ts}$, which is reproduced here:

\begin{figure}[!tb]
 \begin{center}
\resizebox{120mm}{!}{
\includegraphics{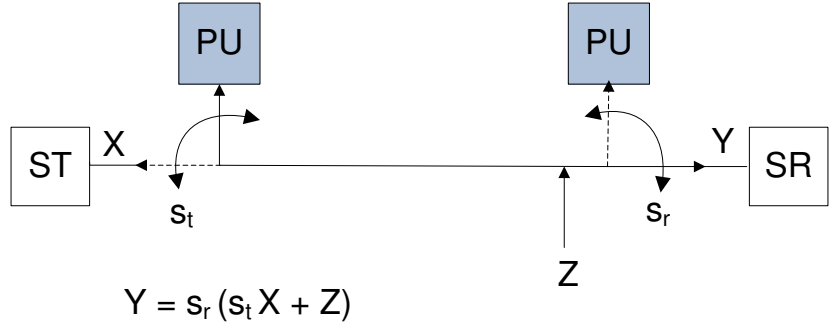}
}
\end{center}
\caption{The two-switch cognitive radio model proposed in \cite{jafar.07b}. } \label{fig:fig2}
\end{figure}

\begin{equation} \label{eq:cts}
C_{ts}\left(P_S\right)={\rm Pr}\left(S_t,S_r=1\right)\log_2\left(1+\frac{P_S}{{\rm
Pr}\left(S_t,S_r=1\right)}\right),
\end{equation} where ${\rm Pr} \left(\theta\right)$ is the probability of a given event $\theta$.

By its turn, the overlay approach is considered to be a form of cooperative transmission by Devroye {\it et al} \cite{devroye.07,devroye.06b}. The authors consider the case where two users asymmetrically cooperate to send two independent messages to two separate non-cooperative receivers, and the case when the second transmitter operates as a relay in order to help the primary transmitter. In both cases, it is assumed perfect channel state information at the primary and secondary transmitters, and that the secondary transmitter has  non-causal knowledge of the message to be transmitted by the primary.

Jovicic and Viswanath \cite{jovicic.06} introduced a computation for the capacity of both transmitters in the overlay cognitive radio model, very similar to that in \cite{devroye.07,devroye.06b}, assuming that there is no change in the encoding and decoding of the primary user. In other words, the primary transmitter and receiver are unaware of the secondary user. Figure \ref{fig:fig3} illustrates the model considered in \cite{jovicic.06}. $PT$ is the primary transmitter, $PR$ is the primary receiver, $X_P$ is the message to be transmitted by the primary, $Y_P$ is the message received at the primary receiver, $Z_P$ is random noise at the primary receiver, $a$ is the channel gain between $ST$ and $PR$ and $b$ is the channel gain between $PT$ and $SR$. The rest of the notation follows the definitions for Figure \ref{fig:fig2}. $X_P$ and $X_S$ are linked in Figure \ref{fig:fig3} to indicate that the secondary transmitter must have {\it a priori} knowledge of the message that will be sent by the primary transmitter. In \cite {jovicic.06} it is shown that, while the primary transmitter is unaffected, the secondary transmitter can achieve a rate as high as:

\begin{figure}[!tb]
 \begin{center}
\resizebox{80mm}{!}{
\includegraphics{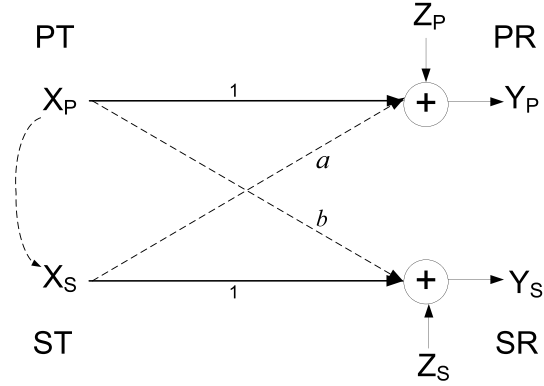}
}
\end{center}
\caption{The overlay cognitive radio model proposed in \cite{jovicic.06}.} \label{fig:fig3}
\end{figure}

\begin{equation} \label{eq:cover}
C_{over} = \frac{1}{2}\log_2\left(1+\left(1-\alpha^*\right)P_S\right),
\end{equation}where
\begin{equation}
\alpha^*=\left(\frac{\sqrt{ P_P}\left(\sqrt{ 1+a^2P_S\left(1+P_P\right)}-1\right)}{a\sqrt{ P_S}\left(1+P_P\right)}\right)^2,
\end{equation}$P_P$ is the primary transmission power, $a \leq 1$ and $\alpha$ means the amount of power the secondary transmitter uses to assist the primary communication and can assume values between zero and one. If $\alpha = 0$, there is no assistance to the primary communication. On the other hand, if $\alpha=1$ the total power of the secondary transmitter is used to assist the primary communication.  The constraint that $a \leq 1$ in the model in \cite{jovicic.06} means that the system is affected by a low interference gain, or that the secondary transmitter is closer to the secondary receiver than to the primary receiver. Note the independence of $b$ in (\ref{eq:cover}). That is due to the fact that the secondary transmitter has {\it a priori} knowledge of what the primary transmitter will send, and also due to the use of dirty paper coding techniques \cite{costa.83}.

Srinivasa and Jafar \cite{jafar.07b} indicate that the secondary transmitter capacity  achieved with the overlay approach can be considerably greater than the secondary transmitter capacity achieved through the interweave approach. The clear advantage of the overlay cognitive radio model is the possibility to have two concurrent transmissions within the same interference region.

We finish this section by noting that it is possible to draw an analogy between the two-switch cognitive radio model in \cite{jafar.07b} and the definition of the interference region. When a node is transmitting the other nodes within its interference region can not transmit, just as if their switches were open in the two-switch model. The transmitting node can be seen as the primary user, while the other nodes in its interference region are seen as secondary transmitters. At this point it is interesting to ask what would happen if the nodes within a given interference region could communicate based on an overlay cognitive model as in \cite{jovicic.06} instead of the two-switch model. While a given node is transmitting (the primary user), another node in its interference region (the secondary user) could transmit to a different node. In the next section we elaborate on this issue.

\section{Cognitive Radio and WMNs} \label{ref:crwmn}

One of the main challenges on the practical application of the overlay model is the fact that the secondary transmitter must know the message that will be sent by the primary transmitter. From an information theoretic perspective, if the secondary transmitter does not have such {\it a priori} information (non-causally), it may be possible to obtain such information causally, if the capacity of the channel between both transmitters is much greater than the capacity of the channel between the primary transmitter and the primary receiver \cite{jafar.07a}. In practice that might be hard to implement, and it would require both transmitters to be relatively close to each other.

However, consider the case of a WMN where an intermediate node is relaying a message to the next node towards the gateway. The source node, or a node that relayed that same message at the beginning of the chain of hops, knows what is going to be transmitted from that intermediate node. Therefore, it may be seen as the secondary transmitter on the overlay cognitive radio model, while the intermediate node would be the primary transmitter. Due to the particular topology of a WMN, and supposing perfect synchronization, {the issue of the non-casual knowledge of the primary transmitter message is naturally overcame by the sequential nature of WMNs transmissions.

Note that usually in cognitive radio applications the primary and the secondary transmitters belong to different communication systems. For instance, the primary may be a broadcast TV transmitter, while the secondary may be part of a wireless local area network. In our scenario both the primary and the secondary transmitters belong to the same communication system, a wireless mesh network. Moreover, the nodes can dynamically operate as either primary or secondary transmitters, depending on the transmission schedule.

In a typical cognitive radio scenario, the primary transmitter is supposed to transmit at least at the same rate as if the secondary transmitter was not there. Here, for fairness purposes, we want both primary and secondary transmitters to have the same rate. For that sake we consider the model defined as the interference channel with degraded message sets (IC-DMS) discussed in \cite{jovicic.06}, which was also analyzed in \cite{wu.07}. The achievable rate regions for that channel, the primary transmission rate $R_P$ and the secondary transmission rate $R_S$, are given by \cite{jovicic.06}:

\begin{equation}
0 \leq R_P \leq \frac{1}{2}\log_2\left(1+\frac{\left(\sqrt{ P_P} + a\sqrt{ \alpha P_S}\right)^2}{1+a^2\left(1-\alpha\right)P_S}\right),
\end{equation}
\begin{equation}
0 \leq R_S \leq \frac{1}{2} \log_2 \left(1+\left(1-\alpha\right)P_S\right).
\end{equation}Then, we search for the value of $\alpha$ that will guarantee the pair of closest rates ($R_P,R_S$) for a given value of $a$, $P_P$ and $P_S$. In this case, and for $a\neq0$, we enforce that the two concurrent transmitters will operate at the rate dictated by $\min(R_P,R_S)$. Thus, the primary achievable rate will be smaller than the rate that the primary user would experience if the secondary user was not active.

In order to measure the advantage of empolying concurrent transmissions within the same interference region according to the overlay cognitive radio model, let $B$ be the original rate of the primary user. Depending on the cross coefficient $a$, different pairs of rates ($R_P,R_S$) can be achieved. Let $\gamma$ be defined as
\begin{equation}
\gamma=\frac{\min\left(R_P,R_S\right)}{B},
\end{equation}where $0<\gamma<1$. Figure \ref{fig:figPpPs} shows $\gamma$ as a function of $a$ for $P_P=P_S \in \{1,2,5,10,20,30,40\}$. From the figure it is clear that $\gamma$ decreases with the increase of the transmission power. This result is justified by the fact that the greater is either the transmission power or the channel gain $a$, the greater is the interference experienced by the primary receiver due to the secondary transmitter. So, the secondary transmitter has to use more power (greater $\alpha$) to assist the primary communication and, therefore, less power will last for its own communication. As a consequence, its rate will be smaller and so $\gamma$, since it is a function of $\min\left(R_P,R_S\right)$.

\begin{figure}[!tb]
 \begin{center}
\resizebox{120mm}{!}{
\includegraphics{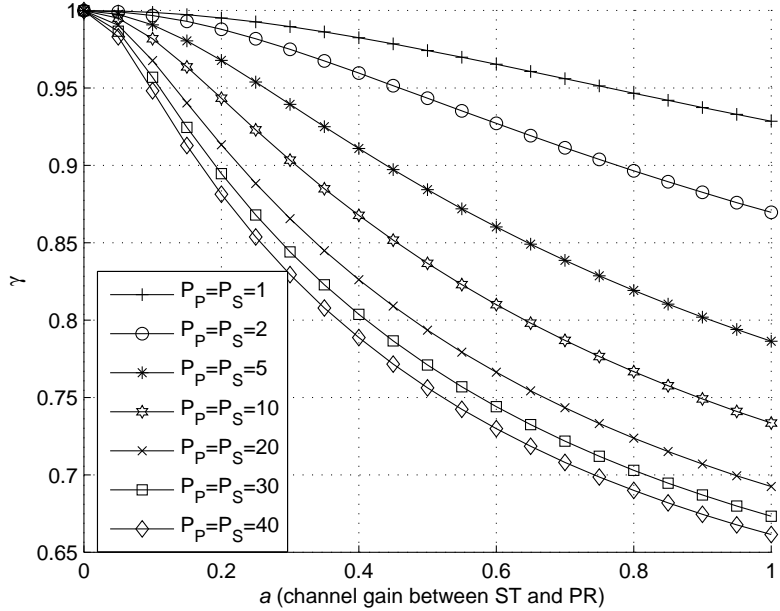}
}
\end{center}
\caption{Factor $\gamma$ as a function of $a$ (channel gain between $ST$ and $PR$) and $P_P=P_S$.} \label{fig:figPpPs}
\end{figure}

In practical scenarios of WMNs, it is fair to assume small values for $a$, since the primary and secondary links must have at least another link between them. Moreover, the further the primary receiver and the secondary transmitter are (smaller $a$), the larger is the capacity of both transmitters. Supposing a log-distance path loss model \cite{rappaport} with an exponent between 3 and 4, that the channel gain between $PT$ and $PR$ is 1, and that the distance between $ST$ and $PR$ is about three times greater than that of $PT$ and $PR$, it is fair to say that $a$ in our case will be small (between 0.1 and 0.2). For this case, even for large values of power, from Figure \ref{fig:figPpPs} we can see that $\gamma$ is at least close to 0.9. In the sequel we consider $a=0.2$ and $P_P=P_S=10$ for evaluating numerical results. In this case, $\gamma\simeq0.95$. Then, two concurrent transmissions with instantaneous rates of $0.95B$ bits per channel use can occur within a certain interference region, where before only one transmission occurred at a instantaneous rate of $B$ bits per channel use. Therefore, we expect to obtain a considerable capacity improvement when applying the overlay cognitive radio model for the communication between nodes in a WMN.

\section{Numerical Results}

In order to investigate the effect of considering that nodes in a WMN can communicate using the overlay cognitive radio model \cite{jovicic.06}, three kinds of WMNs will be discussed. The first one is the simple chain of nodes as introduced earlier. The second one is a generalization of the first one, a set of regular topologies. The last one is an arbitrary topology with the gateway placed at the center and the nodes randomly placed around the gateway. Throughout this section we consider that active nodes generate the same amount of traffic, that each node has an infinite packet buffer and retains copies of the packets it forwards until that packet is transmitted to the gateway or has been forwarded on by the nodes within its interference region, and that $\gamma=0.95$ ($a=0.2$, $P_P=P_S=10$). Moreover, we define $C^{WMN}$ to be the ordinary wireless mesh network capacity, and $C^{WMN}_{over}$ to be the wireless mesh network capacity when using the overlay cognitive radio model for the communication between nodes. Comparisons will be based on the ratio between these two capacities.

\subsection{Simple Chain Topology}

In the first case we consider the same simple WMN composed of a chain of nodes showed in Figure \ref{fig:fig1}. Applying the cognitive radio model, the second transmission from node 8 to node 7 can occur at the same time that node 6 is forwarding to node 5 the previous message sent by node 8. The second transmission from node 8 to node 7 would be seen as the secondary transmission relative to the first transmission from node 6 to 5, which of course would be the primary transmission. Note that in Figure \ref{fig:fig1} the second transmission from node 8 occurs only during the first transmission from node 3 to node 2. Generalizing this example, we have the transmission schedule shown in Figure \ref{fig:fig4}. The transmissions represented by solid arrows are considered as primary transmissions regarding the cognitive radio model. The empty arrows represent the secondary transmissions that can occur in the primary's interference region since the message of the primary transmitter is known by the secondary transmitter.

\begin{figure}[!tb]
 \begin{center}
\resizebox{100mm}{!}{
\includegraphics{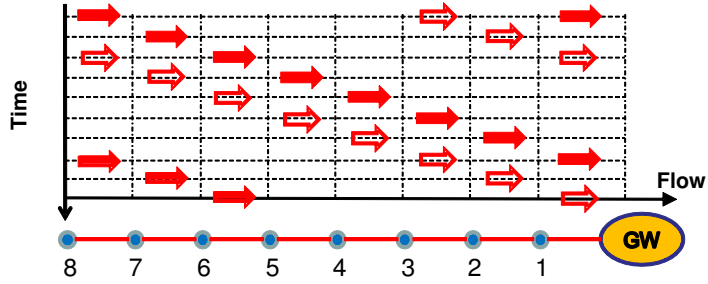}
}
\end{center}
\caption{Example of a WMN composed of a chain of nodes and a gateway. The nodes are able to communicate with each other by means of the overlay cognitive radio model. The transmissions represented by solid arrows are considered as primary transmissions. The empty arrows represent the secondary transmissions.  The interference range is three hops.} \label{fig:fig4}
\end{figure}

Comparing Figures \ref{fig:fig1} and \ref{fig:fig4} we can predict that by using the cognitive radio model the network capacity will be increased. Indeed, taking into account the same definition for the interference range as used in Figure \ref{fig:fig1} (three times the transmission range), we can show that $C^{WMN}_{over}=\gamma 2B/7$. Since we assumed $\gamma=0.95$, this is a 36$\%$ increase because in this case $C^{WMN}=B/5$.

\subsection{Regular Topology}

Consider now a regular topology as shown in Figure \ref{fig:fig8} with three possible different arrangements: nodes $\{33,37\}$ generate traffic, nodes $\{33, 35, 37, 39\}$ generate traffic, and nodes $\{33, 34, 35, 36, 37, 38, 39, 40\}$ generate traffic. Therefore, we have a generalization of the previous setup, where now we consider two chains, four chains and eight chains of nodes. Considering $\gamma=0.95$, we can show that there is a capacity improvement of $19\%$, $22\%$ and $16\%$ for the cases of two, four and eight nodes generating traffic, respectively.

For instance, consider the case where the overlay cognitive radio model is not used and that only nodes 33 and 37 generate traffic towards the gateway. Therefore, we have two different network branches separated by the gateway. When node 13 is transmitting to node 5, node 9 can not transmit to node 1 since nodes 9 and 1 are within the interference region of nodes 13 and 5, even though they are in another network branch. Following this procedure, where each node has to respect the interference regions not only of the nodes within its own chain, but also of the nodes within neighbor chains, we can show that in the steady state each packet from nodes 33 and 37 takes in average 5 transmission instants to reach the gateway. Therefore $C^{WMN}=B/5$. A different schedule is obtained when applying the overlay cognitive radio model. For instance, while node 5 is forwarding the first packet of node 37 to the gateway, node 21 is already forwarding to node 13 the second packet of node 37. Generalizing, we can show that each packet from either nodes 33 or 37 will take in average 4 transmission instants to reach the gateway, which means a capacity of $C^{WMN}_{over}=\gamma B/4$. Therefore, since we assumed $\gamma=0.95$, the improvement is of $19\%$.

In the case of eight nodes generating traffic the gain is smaller since there is an excess of branches close to each other. Many of the cognitive transmission opportunities that appeared in the other cases can not be exploited in this one since they would interfere with the transmission of some other branch. This indicates that the future investigation of inter branches cooperation can bring additional capacity gains. Finally, we end this section by noting that such regular topologies as those in Figure \ref{fig:fig8} are not very likely in practice due, for instance, to topographical constraints and user density.

\begin{figure}[!tb]
 \begin{center}
\resizebox{120mm}{!}{
\includegraphics{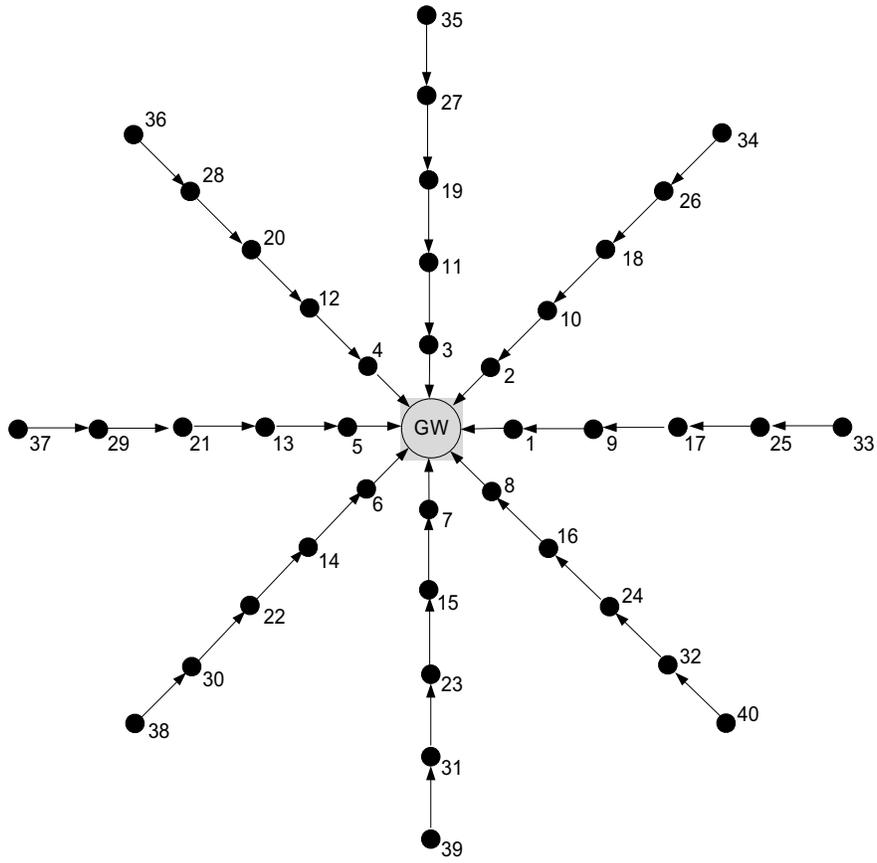}
}
\end{center}
\caption{Regular WMN topology.} \label{fig:fig8}
\end{figure}

\subsection{Arbitrary Topology}

Let us now consider a more general, and practical, network topology shown in Figure \ref{fig:fig5}. The interference range is considered to be three times the transmission range, as in the previous examples. Therefore, each node has a possibly different interference region which has to be taken into account in the capacity calculations. The interference regions are listed in Table \ref{tab:tabela}.

\begin{figure}[!tb]
 \begin{center}
\resizebox{120mm}{!}{
\includegraphics{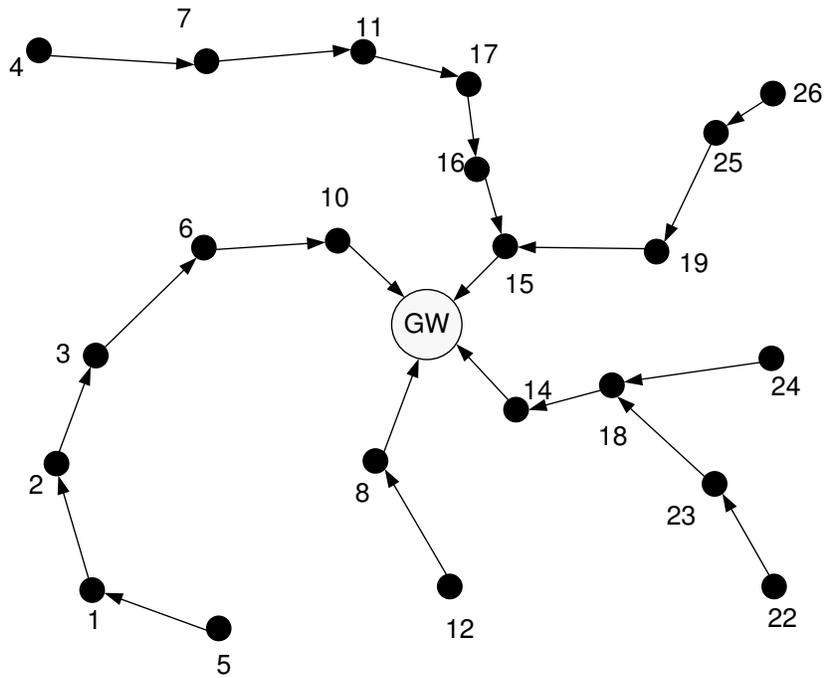}
}
\end{center}
\caption{Arbitray WMN topology.} \label{fig:fig5}
\end{figure}

We consider a scenario where six nodes $\{4, 5, 12, 22, 24, 26\}$ are generating traffic. In this case we can show that by using the overlay cognitive radio model, and supposing $\gamma=0.95$, the capacity improvement is of around 36$\%$ with respect to the ordinary case (capacities of $C^{WMN}_{over}=\gamma B/12$ and $C^{WMN}=B/17$). Furthermore, note also that in the subnetwork composed of nodes 8 and 12 the overlay cognitive radio model is never used. Therefore, the improvement could be even larger if the packets coming from node 12 required one more hop to reach the gateway.

\begin{table}[!tb]
\begin{center}
\begin{tabular}{||c|c||}
\hline \hline
\bf{Node} & \bf{Nodes in its Interference Region} \\ \hline
1 & 2,3,5,8,12\\ \hline
2 & 1,3,5,6,8,10\\ \hline
3 & 1,2,4-8,10,14,15,GW\\ \hline
4 & 3,6,7,10,11,17\\ \hline
5 & 1-3,8,12,14,GW\\ \hline
6 & 2-4,7,8,10,11,15-17,GW\\ \hline
7 & 3,4,6,10,11,15-17,GW\\ \hline
8 & 1-3,5,6,10,12,14-16,18,19,22-24,GW\\ \hline
10 & 2-4,6-8,11,14-19,25,GW\\ \hline
11 & 4,6,7,10,15-17,19,25,26,GW\\ \hline
12 & 1,5,8,14,15,18,22-24,GW \\ \hline
14 & 3,5,6,8,10,12,15-19,22-25,GW\\ \hline
15 & 3,6-8,10-12,14,16-19,23-26,GW\\ \hline
16 & 6-8,10,11,14,15,17-19,24-26,GW\\ \hline
17 & 4,6,7,10,11,14-16,18,19,25,26,GW\\ \hline
18 & 8,10,12,14-17,19,22-26,GW\\ \hline
19 & 8,10,14-18,23-26,GW\\ \hline
22 & 8,12,14,18,23,24\\ \hline
23 & 8,12,14,15,18,19,22,24,GW\\ \hline
24 & 8,12,14-16,18,19,22,23,25,26,GW\\ \hline
25 & 10,11,14-19,24,26,GW\\ \hline
26 & 11, 15-19,24,25\\ \hline
GW & 3,5-8,10-12,14-19,23-25\\ \hline
\end{tabular}
\end{center}
\caption{Inteference region for the nodes involved in the calculations of the capacity of the network shown in Figure \ref{fig:fig5}.}
\label{tab:tabela}
\end{table}

\begin{table}[!tb]
\begin{center}
\begin{tabular}{||c|c||}
\hline \hline
{\bf Topology} & {\bf Improvement} \\ \hline
Simple Chain & 36$\%$ \\ \hline
Regular - 2 branches & 19$\%$ \\ \hline
Regular - 4 branches & 22$\%$ \\ \hline
Regular - 8 branches & 16$\%$ \\ \hline
Arbitrary & 36$\%$ \\ \hline \hline
\end{tabular}
\end{center}
\caption{Capacity improvement achieved by utilizing the overlay cognitive radio model for the communication between nodes in a WMN.}
\label{tab:tab2}
\end{table}

\section{Final Comments} \label{sec:conc}

WMNs and cognitive radios are two subjects of intense research nowadays. We propose a way to bring these two ideas together. Our results show that the capacity of a WMN can be substantially increased by considering that the network nodes can communicate using the overlay cognitive radio model. Such a model allows for two concurrent transmissions within a given interference region.

Table \ref{tab:tab2} summarizes the results obtained in this paper, in terms of the capacity improvement achieved by utilizing the overlay cognitive radio model. From the table we can see that for the practical case of an arbitrary (randomly generated) topology the capacity gain was of $36\%$.

There are many open questions related to our proposal. For instance, how does the capacity gain behave with nodes density, number of gateways, type of traffic? How to synchronize the primary and secondary transmitters? How to design a real transmission scheme that can approach the predicted performance of the overlay cognitive radio model? The promising results motivate further investigations in the potentials and implications of the current proposal.

\end{document}